\documentclass[aps,prc,twocolumn,superscriptaddress,floatfix]{revtex4-1}
\usepackage[english]{babel}
\usepackage[T1]{fontenc}
\usepackage[colorlinks=true,linkcolor=blue,citecolor=blue,urlcolor=blue]{hyperref}
\usepackage{bm,bbm,amssymb,amsmath}
\usepackage{graphicx}
\usepackage{multirow,booktabs}
\usepackage{isotope}
\usepackage{physics}
\usepackage[capitalize]{cleveref}

\begin{document}

\title{Single- and two-nucleon momentum distributions for local chiral interactions}

\author{D. Lonardoni}
\affiliation{Facility for Rare Isotope Beams, Michigan State University, East Lansing, Michigan 48824, USA}
\affiliation{Theoretical Division, Los Alamos National Laboratory, Los Alamos, New Mexico 87545, USA}

\author{S. Gandolfi}
\affiliation{Theoretical Division, Los Alamos National Laboratory, Los Alamos, New Mexico 87545, USA}

\author{X.~B. Wang}
\affiliation{School of Science, Huzhou University, Huzhou 313000, China}

\author{J. Carlson}
\affiliation{Theoretical Division, Los Alamos National Laboratory, Los Alamos, New Mexico 87545, USA}

\begin{abstract}
We present quantum Monte Carlo calculations of the single- and two-nucleon 
momentum distributions in selected nuclei for $A\le16$.
We employ local chiral interactions at next-to-next-to-leading order.
We find good agreement at low momentum with the single-nucleon momentum 
distributions derived for phenomenological potentials.
The same agreement is found for the integrated two-nucleon momentum distributions 
at low relative momentum $q$ and low center-of-mass momentum $Q$. 
We provide results for the two-nucleon momentum distributions as a function 
of both $q$ and $Q$. The large ratio
of $pn$ to $pp$ pairs around $q=2\,\rm fm^{-1}$ for back-to-back $(Q=0)$ pairs
is confirmed up to \isotope[16]{O}, and results are compatible with 
those extracted from available experimental data.
\end{abstract}

\maketitle

\section{Introduction}
Quantum Monte Carlo (QMC) methods have been extensively used in the past 
to derive properties of strongly correlated systems, 
including nuclei, neutron drops, and neutron and nuclear matter 
(see Ref.~\cite{Carlson:2015} for a recent review).
Part of their success relies on the possibility to tackle the 
nuclear many-body problem in a nonperturbative fashion, by employing 
accurate wave functions that include two- and three-body correlations.

Momentum distributions of individual nucleons and nucleon pairs
strongly depend on such correlations, as they reflect 
features of the short-range structure of nuclei. 
While these momentum distributions are not directly observable,
because they are coupled with, e.g., electromagnetic current operators,
they do provide a strong influence on some observables such as back-to-back
nucleons measured in quasielastic scattering.
For instance, it was found that the strong spatial-spin-isospin correlations 
induced by the tensor force lead to large differences in the $pp$ and $pn$ 
distributions at moderate values of the relative momentum in the 
pair~\cite{Schiavilla:2007,Alvioli:2008}. 
These differences have been observed in $(e,e'pN)$ experiments on 
\isotope[12]{C} at low momentum~\cite{Subedi:2008} and on \isotope[4]{He} 
at higher momentum~\cite{Korover:2014} at Jefferson Laboratory (JLab).
The same conclusions have been found in heavier systems, including \isotope[27]{Al}, 
\isotope[56]{Fe}, and \isotope[208]{Pb}~\cite{Hen:2014}.

The variational Monte Carlo (VMC) method has been used to 
calculate the momentum distributions in $A\leq12$ 
nuclei~\cite{Schiavilla:1986,Schiavilla:2007,Wiringa:2008,Wiringa:2014,Wiringa:rhok,Wiringa:rhok12} 
by employing phenomenological nuclear interactions, 
i.e., Argonne $v_{18}$ (AV18) nucleon-nucleon $(N\!N)$ potential combined 
with Urbana models (UIX-UX) for the three-nucleon $(3N)$ force~\cite{Carlson:2015}.
The same family of potentials has been employed in 
cluster expansion methods, including the cluster VMC algorithm~\cite{Lonardoni:2017},
to calculate the momentum distributions of heavier systems, such as \isotope[16]{O} and 
\isotope[40]{Ca}~\cite{Alvioli:2005,Alvioli:2008,Alvioli:2013,Alvioli:2016,Lonardoni:2017}.

In this work, we present QMC calculations of single- and two-nucleon momentum
distributions in \isotope[4]{He}, \isotope[12]{C}, and \isotope[16]{O} employing
the local chiral effective field theory (EFT) interactions at next-to-next-to leading order 
(N$^2$LO) developed in 
Refs.~\cite{Gezerlis:2013,Gezerlis:2014,Tews:2016,Lynn:2016,Lonardoni:2018prc}.

\section{Hamiltonian and wave function}
Nuclei are described as a collection of point-like particles of mass $m$ interacting 
via two- and three-body potentials according to the nonrelativistic Hamiltonian
\begin{align}
	H=-\frac{\hbar^2}{2m}\sum_i \nabla_i^2+\sum_{i<j}v_{ij}+\sum_{i<j<k}V_{ijk}.
\end{align}
In this work we consider the local chiral interactions at N$^2$LO of 
Refs.~\cite{Gezerlis:2013,Gezerlis:2014,Tews:2016,Lynn:2016}.

The long-range part of the $N\!N$ potential is given by pion-exchange contributions
that are determined by the chiral symmetry of quantum chromodynamics and 
low-energy pion-nucleon scattering data. 
The short-range terms are given by contact interactions, described by low-energy 
constants (LECs) that are fit to nucleon-nucleon scattering data~\cite{Gezerlis:2014}.
At N$^2$LO, the two-body local chiral potential is written as a sum of radial 
functions multiplying spin and isospin operators, which correspond to
the first seven terms of the AV18 potential, i.e.,
$\mathcal O_{ij}^{p=1,7} = \big[\mathbbm 1,\bm\tau_i\cdot\bm\tau_j,
\bm\sigma_i\cdot\bm\sigma_j,\bm\sigma_i\cdot\bm\sigma_j\,\bm\tau_i\cdot\bm\tau_j,
S_{ij},S_{ij}\,\bm\tau_i\cdot\bm\tau_j,\vb{L}\cdot\vb{S}\big]$,
where $S_{ij}$ is the tensor operator, and $\vb{L}$ and $\vb{S}$ are
the relative angular momentum and the total spin of the nucleon pair $ij$, respectively.

The $3N$ local chiral interaction at N$^2$LO is written as a sum of two-pion
exchange (TPE) contributions plus shorter-range terms, $V_D$ and $V_E$. 
The LECs of the TPE terms are the same as those of the two-body sector, while the
additional LECs for the shorter-range terms are fit to few-body observables.
In more detail, $c_D$ and $c_E$ are fit to the binding energy of 
\isotope[4]{He} and $n$-$\alpha$ scattering $P$-wave phase shifts,
providing a probe to the properties of light nuclei, spin-orbit
splitting, and $T=3/2$ physics~\cite{Lynn:2016}.
According to the Fierz-rearrangement freedom, different equivalent operator 
structures are possible for locally regularized three-body contact operators 
at N$^2$LO~\cite{Epelbaum:2002}. We employ here the $E\tau$ and $E\mathbbm1$ 
parametrizations for $V_E$, corresponding to the choice of the $\bm\tau_i\cdot\bm\tau_j$ isospin operator
and the identity operator $\mathbbm1$, respectively. We use the coordinate-space cutoffs 
$R_0=1.0\,\rm fm$ and $R_0=1.2\,\rm fm$, which correspond roughly to cutoffs in momentum space of 
$500$ and $400\,\rm MeV$~\cite{Lynn:2017,Hoppe:2017}, respectively. 
As shown in Refs.~\cite{Lynn:2017,Lonardoni:2018prl,Lonardoni:2018prc}, the use of different 
three-body operator structures and coordinate-space cutoffs leads to  
very similar ground-state properties in light and medium-mass nuclei.

We perform QMC calculations of the single- and two-nucleon momentum distributions
by employing the trial wave function used in auxiliary field diffusion Monte Carlo
(AFDMC) calculations of light- and medium-mass nuclei~\cite{Lonardoni:2018prl,Lonardoni:2018prc}.
Such a wave function takes the form
\begin{align}
&\langle RS|\Psi\rangle=\;\langle RS|\prod_{i<j}f^1_{ij}\,\prod_{i<j<k}f^{3c}_{ijk} \nonumber \\ 
&\times\left[\mathbbm1+\sum_{i<j}\sum_{p=2}^6 f^p_{ij}\,\mathcal O_{ij}^p\, f_{ij}^{3p}+\sum_{i<j<k}U_{ijk}\right]|\Phi\rangle_{J^\pi,T} ,
\label{eq:psi}
\end{align}
where $|RS\rangle$ are the $3A$ spatial coordinates and $4A$ spin/isospin amplitudes for each nucleon.
The pair correlation functions $f^p_{ij}$ are obtained as the solution
of Schr\"odinger-like equations in the relative distance between two particles, as explained in Ref.~\cite{Carlson:2015}. 
$f^{3c}_{ijk}$ and $f^{3p}_{ij}$ are spin/isospin-independent functions introduced to reduce the strength of the 
spin/isospin-dependent correlations when other particles are nearby~\cite{Pudliner:1997}.
$U_{ijk}$ are three-body spin/isospin-dependent correlations, whose operator structure resembles that
of the $3N$ potential $V_{ijk}$. 
The term $|\Phi\rangle$ represents the mean-field part of the wave function. 
It consists of a sum of Slater determinants $\mathcal D$ constructed using shell-model-like single-particle orbitals
\begin{align} 
\langle RS|\Phi\rangle_{J^\pi,T} = \sum_n c_n\Big[\sum \mathcal C_{J\!M}\,\mathcal D\big\{\phi_\alpha(\vb{r}_i,s_i)\big\}_{J,M}\Big]_{J^\pi,T} ,
\label{eq:phi}
\end{align}
where $\vb{r}_i$ are the spatial coordinates of the nucleons, and $s_i$ represent their spinors.
The Clebsch-Gordan coefficients $\mathcal C_{J\!M}$ are chosen to reproduce the experimental 
total angular momentum, total isospin, and parity $(J^\pi,T)$ of the nucleus, while the
$c_n$ are variational parameters multiplying different wave-function components having the same quantum numbers. 
Each single-particle orbital $\phi_\alpha$ consists of a radial function, bound-state solution of a 
Woods-Saxon wine-bottle potential, multiplied by the proper spherical harmonic and the spin/isospin state.
For closed-shell systems, such as \isotope[4]{He} and \isotope[16]{O}, the mean-field wave function
of~\cref{eq:phi} is given by a single Slater determinant. In \isotope[12]{C}, 119 determinants constructed with
$p$-shell single-particle orbitals need to be coupled in order to obtain a $(0^+,0)$ state
with good binding energy~\cite{Lonardoni:2018prc}. 
However, observables like the charge radius are well determined by using a reduced subset of Slater determinants.
A trial wave function including only 13 Slater determinants provides a $(0^+,0)$ state with the same charge radius 
as the full $p$-shell wave function, even though the total VMC energy is reduced by $\approx 3\,\rm MeV$.
Such a simplified wave function has been used in this work for the VMC estimate of single- and two-nucleon momentum distributions 
in \isotope[12]{C}, a calculation otherwise computationally prohibitive.
Details on the construction of the wave functions can be found in Ref.~\cite{Lonardoni:2018prc}.

According to the VMC method, given the trial wave function $\Psi_T=\langle RS|\Psi\rangle_{J^\pi,T}$,
the expectation value of the Hamiltonian is given by
\begin{align}
E_V=\langle H\rangle=\frac{\langle\Psi_T|H|\Psi_T\rangle}{\langle\Psi_T|\Psi_T\rangle}\geq E_0 ,
\label{eq:ev}
\end{align}
where $E_0$ is the energy of the true ground state with the same quantum numbers as $\Psi_T$. 
The equality in the above equation is only valid if the wave function is the exact ground-state 
wave function $\Psi_0$; i.e., the variational energy is always an upper bound to the 
true ground-state energy. $E_V$ depends in general on the employed wave function.
By minimizing the energy expectation value of~\cref{eq:ev} with respect to changes 
in the variational parameters of $\Psi_T$, one obtains an optimized wave function, 
i.e., the best approximation of $\Psi_0$, which can be used to calculate other quantities 
of interest, such as the momentum distributions. We optimize our trial wave functions for
local chiral interactions at N$^2$LO. 
During the optimization a constraint is used in order to 
approximatively obtain the experimental charge radii, which are reported in~\cref{tab:rch}.
Note that these are VMC results only, while the charge radii of Ref.~\cite{Lonardoni:2018prc}
correspond to the extrapolated results from mixed estimates: 
$2\,\langle r_{\rm ch}^{\rm AFDMC}\rangle-\langle r_{\rm ch}^{\rm VMC}\rangle$.
Differences between extrapolated and VMC results are, however, within statistical uncertainties.
The true ground state of the system can finally be obtained by using the AFDMC method.
The imaginary time propagation is used to project out the lowest-energy state with the symmetry 
of the trial wave function $\Psi_T$:
\begin{align}
	\Psi_0=\lim_{\tau\to\infty}e^{-(H-E_T)\tau}\,\Psi_T,
\end{align} 
where $E_T$ is a parameter that controls the normalization (see Ref.~\cite{Lonardoni:2018prc}
for more details). Although the imaginary time propagation allows one
to access properties of the true ground state of the system, the AFDMC calculation of 
two-nucleon momentum distributions is at present computationally prohibitive.  
For this reason, in this work we present VMC results only, providing an example of the AFDMC 
calculation for single-nucleon momentum distribution in Sec.~\ref{sec:rhok}.

\begin{table}[htb]
\centering
\caption[]{VMC charge radii (in fm) for the optimized wave function of~\cref{eq:psi}
and different N$^2$LO local chiral potentials. Experimental results are also shown.}
\begin{ruledtabular}
\begin{tabular}{lccc}
Nucleus & $V_E,\,R_0\,(\rm fm)$ & VMC & Expt. \\[0.1cm]
\hline
\isotope[4]{He}\,$(0^+,0)$ & $E\tau,\,1.0$      & $1.67(1)$ & $1.680(4)$~\cite{Sick:2008}  \\
                           & $E\mathbbm1,\,1.2$ & $1.64(1)$ &                              \\ [0.2cm]
\isotope[12]{C}\,$(0^+,0)$ & $E\tau,\,1.0$      & $2.48(2)$ & $2.471(6)$~\cite{Sick:1982}  \\ [0.2cm]
\isotope[16]{O}\,$(0^+,0)$ & $E\tau,\,1.0$      & $2.77(3)$ & $2.730(25)$~\cite{Sick:1970} \\
                           & $E\mathbbm1,\,1.2$ & $2.57(3)$ &                              \\
\end{tabular}
\end{ruledtabular}
\label{tab:rch}
\end{table}

\section{Single- and two-nucleon momentum distributions}
The probability of finding a nucleon with momentum $k$ in a given isospin state is proportional to the density 
\begin{align}
\rho_{N}(\vb{k})=&\frac{1}{A}\sum_i\int 
d\vb{r}_1^{\phantom{\prime}}\cdots d\vb{r}_i^{\prime}\,d\vb{r}_i^{\phantom{\prime}}\cdots d\vb{r}_A^{\phantom{\prime}} \nonumber \\
& \times\Psi^\dagger(d\vb{r}_1^{\phantom{\prime}},\ldots, d\vb{r}_i^{\prime},\ldots, d\vb{r}_A^{\phantom{\prime}}) 
  \,e^{-i\vb{k}\cdot(\vb{r}_i^{\phantom{\prime}}-\vb{r}_i^\prime)} \nonumber \\
&\times \mathcal P_N(i)\,\Psi(d\vb{r}_1^{\phantom{\prime}},\ldots, d\vb{r}_i^{\phantom{\prime}},\ldots, d\vb{r}_A^{\phantom{\prime}}), 
\label{eq:rhok}
\end{align}
where 
\begin{align}
\mathcal P_N(i)	=\frac{1\pm\tau_i^z}{2}
\end{align}
is the isospin projection operator for the nucleon $i$,
and $\Psi$ is the optimized wave function of~\cref{eq:psi}.
The normalization is
\begin{align}
\mathcal N_N=\int \frac{d\vb{k}}{(2\pi)^3}\,\rho_N(\vb{k}) ,
\end{align}
where $\mathcal N_N$ is the number of protons or neutrons.

The Fourier transform of~\cref{eq:rhok} can be computed by following a Metropolis Monte Carlo walk 
in the $d\vb{r}_1\ldots d\vb{r}_A$ space and one extra Gaussian integration over $d\vb{r}_i'$ 
at each Monte Carlo configuration, as done in early VMC calculations of few-nucleon momentum 
distributions~\cite{Schiavilla:1986}. It is however convenient to rewrite~\cref{eq:rhok} as
\begin{align}
\rho_N(\vb{k})= 
& \frac{1}{A}\sum_i\int d\vb{r}_1\cdots d\vb{r}_i\cdots d\vb{r}_A \int d\Omega_x \int_0^{x_{\max}}\!\!x^2 dx \nonumber \\
& \times\Psi^\dagger(\vb{r}_1,\ldots,\vb{r}_i+\frac{\vb{x}}{2},\ldots,\vb{r}_A)\,e^{-i\vb{k}\cdot\vb{x}} \nonumber \\
& \times\mathcal P_N(i)\,\Psi(\vb{r}_1,\ldots,\vb{r}_i-\frac{\vb{x}}{2},\ldots,\vb{r}_A) .
\label{eq:rhok2}
\end{align}
In the above equation, the position $\vb{r}_i$ is symmetrically shifted by $\vb{x}/2$ in both left- and 
right-hand wave functions, instead of simply moving the position $\vb{r}_i'$ in the left-hand wave function 
with respect to a fixed position $\vb{r}_i$ in the right-hand wave function. A Gaussian integration is 
performed over $\vb{x}$ by choosing a grid of Gauss-Legendre points $\vb{x}_i$ and sampling the polar 
angle $d\Omega_x$, with a randomly chosen direction for each particle in each Monte Carlo configuration.
This procedure has the advantage of drastically reducing the large statistical errors originating 
from the rapidly oscillating nature of the integrand for large values of $k$~\cite{Wiringa:2014}.
For the systems considered in this work, we obtain good statistics up to $k=10\,\rm fm^{-1}$
integrating to $x_{\max}=12\,\rm fm$ using $120$ Gauss-Legendre points.

Note that the procedure described above cannot be applied in AFDMC calculations, where 
left- and right-hand wave functions are different (see Ref.~\cite{Lonardoni:2018prc} for details).
In this case, \cref{eq:rhok} must be used, and a significant computational effort is needed to achieve 
statistical errors comparable to the corresponding VMC calculation. An example of an AFDMC calculation
of single-nucleon momentum distribution is shown in~\cref{fig:nofk_afdmc}. 

The probability of finding two nucleons in a nucleus with relative momentum
$\vb{q}=(\vb{k}_1-\vb{k}_2)/2$ and total center-of-mass momentum
$\vb{Q}=\vb{k}_1+\vb{k}_2$ in a given isospin state is given by
\begin{align}
\rho_{N\!N}(\vb{q},\vb{Q}) =
& \frac{2}{A(A-1)}\sum_{ij}\int\! 
d\vb{r}_1^{\phantom{\prime}}\!\cdots d\vb{r}_i^{\prime}\,d\vb{r}_i^{\phantom{\prime}} 
d\vb{r}_j^{\prime}\,d\vb{r}_j^{\phantom{\prime}}\cdots d\vb{r}_A^{\phantom{\prime}} \nonumber \\
& \times \Psi^\dagger(d\vb{r}_1^{\phantom{\prime}},\ldots, d\vb{r}_i^{\prime},
d\vb{r}_j^{\prime},\ldots, d\vb{r}_A^{\phantom{\prime}}) \nonumber \\
& \times e^{-i\vb{q}\cdot(\vb{r}_{ij}^{\phantom{\prime}}-\vb{r}_{ij}^\prime)}\,e^{-i\vb{Q}\cdot(\vb{R}_{ij}^{\phantom{\prime}}-\vb{R}_{ij}^\prime)} \nonumber \\
& \times \mathcal P_{N\!N}(ij)\,\Psi(d\vb{r}_1^{\phantom{\prime}},\ldots, d\vb{r}_i^{\phantom{\prime}},
d\vb{r}_j^{\phantom{\prime}},\ldots, d\vb{r}_A^{\phantom{\prime}}) ,
\label{eq:rhoqq}
\end{align}
where $\vb{r}_{ij}=\vb{r}_i-\vb{r}_j$, $\vb{R}_{ij}=(\vb{r}_i+\vb{r}_j)/2$,
and $\mathcal P_{N\!N}(ij)$ is the isospin projector operator for the nucleon pair $ij$:
\begin{align}
\mathcal P_{N\!N}(ij)=\frac{1\pm\tau_i^z}{2}\,\frac{1\pm\tau_j^z}{2} .
\end{align}
The normalization is
\begin{align}
\mathcal N_{N\!N}=\int\frac{d\vb{q}}{(2\pi)^3}\frac{d\vb{Q}}{(2\pi)^3}\,\rho_{N\!N}(\vb{q},\vb{Q}) ,
\end{align}
where $\mathcal N_{N\!N}$ is the number of $pp$, $pn$, or $nn$ nucleon pairs.
Note that integrating $\rho_{N\!N}(\vb{q},\vb{Q})$ over $\vb{Q}$ only gives the probability of 
finding two nucleons with relative momentum $\vb{q}$ regardless their center-of-mass 
momentum $\vb{Q}$, and vice versa.

The integral of~\cref{eq:rhoqq} can be evaluated in a similar fashion to that of~\cref{eq:rhok2}
\begin{align}
\rho_{N\!N}(\vb{q},\vb{Q})= 
& \frac{2}{A(A-1)}\sum_{ij}\int d\vb{r}_1\cdots d\vb{r}_i\, d\vb{r}_j\cdots d\vb{r}_A \nonumber \\
& \times \int d\Omega_x \int_0^{x_{\max}}\!\!x^2 dx \int d\Omega_X \int_0^{X_{\max}}\!\!X^2 dX \nonumber \\
& \times \Psi^\dagger(\vb{r}_1,\ldots,\vb{r}_{ij}+\frac{\vb{x}}{2},\vb{R}_{ij}+\frac{\vb{X}}{2},\ldots,\vb{r}_A) \nonumber \\
& \times e^{-i\vb{q}\cdot\vb{x}}\,e^{-i\vb{Q}\cdot\vb{X}}\,\mathcal P_{N\!N}(ij) \nonumber \\
& \times \Psi(\vb{r}_1,\ldots,\vb{r}_{ij}-\frac{\vb{x}}{2},\vb{R}_{ij}-\frac{\vb{X}}{2},\ldots,\vb{r}_A) ,
\label{eq:rhoqq2}
\end{align}
where now a double Gauss-Legendre integration for each nucleon pair in each Monte Carlo configuration
must be evaluated. This makes the two-nucleon momentum distribution much more computationally 
expensive than the single-nucleon momentum distribution. For the nuclei considered 
in this work we obtain good statistics up to $q=5\,\rm fm^{-1}$ and $Q=3\,\rm fm^{-1}$ integrating 
$x$ to $x_{\max}=12\,\rm fm$ using $120$ Gauss-Legendre points, and $X$ to $X_{\max}=8\,\rm fm$ using 
$80$ Gauss-Legendre points. Note that, because the employed wave functions are eigenstates of the total 
isospin $T$, small effects due to isospin-symmetry-breaking interactions are ignored. 
$T=0$ in all nuclei considered in this work, so it follows that, for a given system, $pp$, $nn$, and 
$T=1$ $pn$ momentum distributions are identical.

\section{Results: single-nucleon momentum distributions}
\label{sec:rhok}

\begin{figure}[b]
	\includegraphics[width=\linewidth]{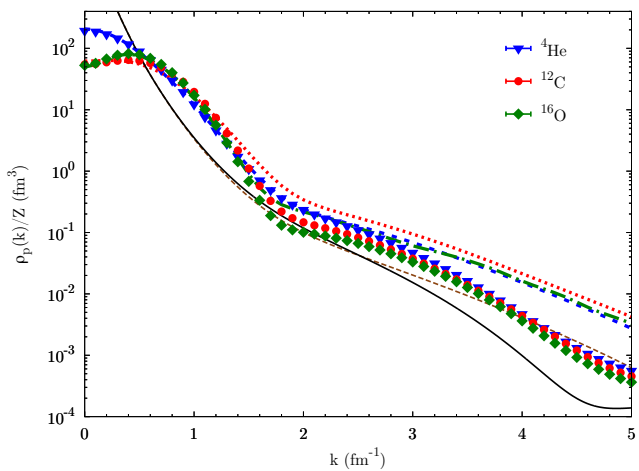}
	\caption[]{Proton momentum distribution in \isotope[4]{He}, \isotope[12]{C}, and \isotope[16]{O}.
		Solid symbols are the results for the N$^2$LO $E\tau$ potential with cutoff $R_0=1.0\,\rm fm$.
		Discontinuous lines are the VMC results for \isotope[4]{He} and \isotope[12]{C}~\cite{Wiringa:2014,Wiringa:rhok}
		and cluster VMC results for \isotope[16]{O}~\cite{Lonardoni:2017} employing the AV18+UIX potential.
		Dashed brown (solid black) line is the deuteron result for AV18~\cite{Wiringa:2014,Wiringa:rhok} 
		(N$^2$LO with cutoff $R_0=1.0\,\rm fm$~\cite{Lynn:2017}).}
	\label{fig:nofk}
\end{figure}

\begin{figure}[t]
	\includegraphics[width=\linewidth]{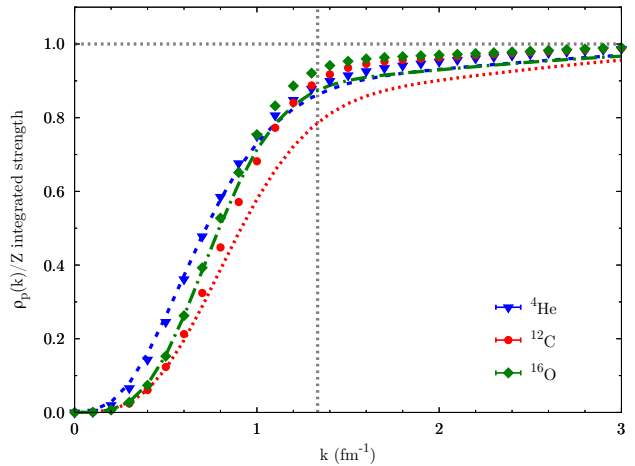}
	\caption[]{Integrated strength of the proton momentum distributions of~\cref{fig:nofk} 
		(the same legend is used). The vertical line indicates the Fermi momentum $k_F$.}
	\label{fig:str}
\end{figure}

The proton momentum distributions $\rho_p(k)$ normalized to the proton number $Z$ 
for the N$^2$LO $E\tau$ potential with cutoff $R_0=1.0\,\rm fm$ are reported in~\cref{fig:nofk}. 
The VMC and cluster VMC results for the AV18+UIX potential of 
Refs.~\cite{Wiringa:2014,Wiringa:rhok,Lonardoni:2017} are also shown for comparison. 
Up to $\approx1.0-1.5\,\rm fm^{-1}$, local chiral interactions
and phenomenological potentials provide a similar description of the proton momentum
distributions. Differences appear at higher momentum, as one would expect being the chiral potentials
derived from a low-energy EFT of the nuclear force. This is also evident by looking at~\cref{fig:str},
where the integrated strength of the proton momentum distribution is shown as a function of~$k$.
At low momentum, chiral and phenomenological results are similar for all nuclei.
At $2\,\rm fm^{-1}$, most of the strength for chiral interactions is already accounted for:
$95.1(1)\%$ in \isotope[4]{He}, $96.3(4)\%$ in \isotope[12]{C}, and $97.0(4)\%$ in \isotope[16]{O}.
At $\approx3.5\,\rm fm^{-1}$ all the strength for chiral interactions is saturated, while phenomenological
potentials still contribute until $\approx4.5-5.0\,\rm fm^{-1}$, as indicated by the higher tail
of $\rho_N(k)$ at high momentum (\cref{fig:nofk}). 
The kinetic energy derived from the single-nucleon momentum distribution,
\begin{align}
K_N=-\frac{\hbar^2}{2m} \int\frac{d\vb{k}}{(2\pi)^3}\,k^2\,\rho_N(\vb{k}) ,	
\end{align}
in general saturates at higher momentum. For both local chiral interactions and phenomenological
potentials, $K_N$ is consistent with the direct VMC calculation for $k\gtrsim6\,\rm fm^{-1}$.
Local chiral interactions result however in $\approx20\%$ to $\approx35\%$ less kinetic energy than 
the phenomenological counterparts.

It is interesting to observe that at high momentum, the tail of the momentum distribution manifests the expected universal 
behavior, i.e., the independence of the high-momentum component of $\rho_N(k)$ upon the specific nucleus.
Such universality has been discussed at length in a number of works (see, for instance, 
Refs.~\cite{Feldmeier:2011,Alvioli:2012,Alvioli:2013,Alvioli:2016}). 
We show here (see~\cref{fig:nofk}) that, depending on the choice of the potential, the universal behavior itself is different.
This is a consequence of the nature of the high-momentum components of the momentum distribution, 
which are determined by short-range correlations, i.e., by the short-range structure of the employed Hamiltonian.
Local chiral interactions and phenomenological potentials are characterized by different short-range physics,
which are reflected in different tails of the momentum distribution.

\begin{figure}[tb]
	\includegraphics[width=\linewidth]{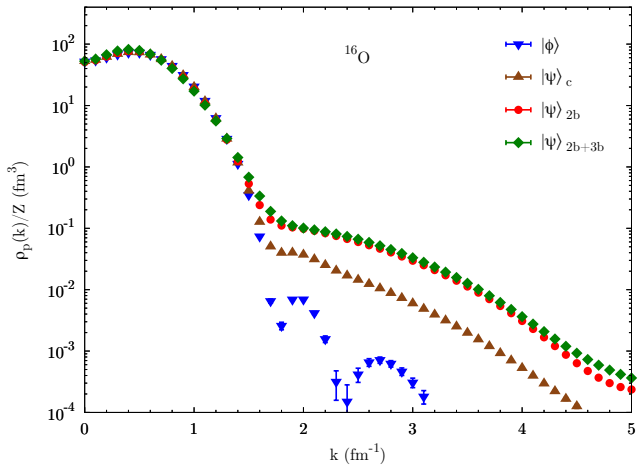}
	\caption[]{Proton momentum distribution in \isotope[16]{O}. 
		$|\Phi\rangle$ is the result for the mean-field wave function of~\cref{eq:phi}.
		$|\Psi\rangle_{\rm c}$, $|\Psi\rangle_{\rm 2b}$, and $|\Psi\rangle_{\rm 2b+3b}$ are the results 
		for the correlated wave function of~\cref{eq:psi} employing spin/isospin-independent two-body, full two-body, 
		and two- plus three-body correlations, respectively.}
	\label{fig:nofk_corr}
\end{figure}

We show in~\cref{fig:nofk_corr} the effect of correlations to the proton momentum distribution
in \isotope[16]{O}. Blue   down triangles refer to the calculation employing the mean-field wave function
of~\cref{eq:phi}. Brown up triangles, red circles, and green diamonds are results for the correlated 
wave function of~\cref{eq:psi} including spin/isospin-independent two-body, full two-body, and two- plus three-body correlations, 
respectively. Results have been obtained by optimizing the different wave functions so as to
obtain the same charge radius reported in~\cref{tab:rch}.
Similarly to the case of phenomenological potentials~\cite{Pieper:1992,Alvioli:2005}, the mean-field part of the
wave function dominates the momentum distribution for $k\lesssim 1.3\,{\rm fm^{-1}}\approx k_F$.
Correlations are fundamental for the construction of higher-momentum components of $\rho_N(k)$, dominated, 
in particular, by two-body spin/isospin correlations. Three-body correlations have a small effect on the 
momentum distribution for chiral interactions, enhancing $\rho_N(k)$ around $2\,\rm fm^{-1}$ and at 
higher momentum, $k>4\,\rm fm^{-1}$.

\begin{figure}[b]
	\includegraphics[width=\linewidth]{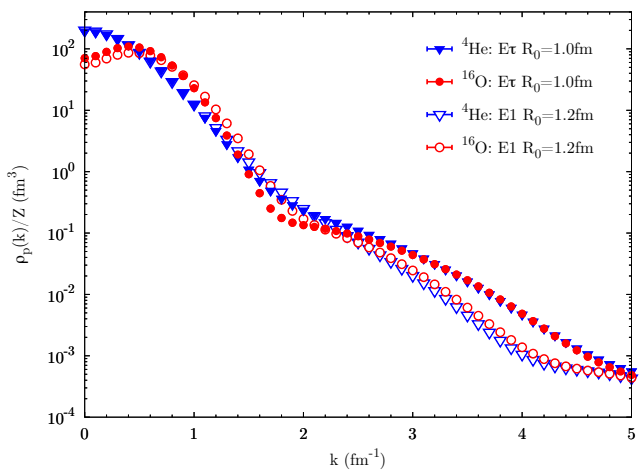}
	\caption[]{Proton momentum distribution in \isotope[4]{He} and \isotope[16]{O}.
		Solid symbols are the results for the N$^2$LO $E\tau$ potential with cutoff $R_0=1.0\,\rm fm$.
		Empty symbols are the results for the N$^2$LO $E\mathbbm1$ potential with cutoff $R_0=1.2\,\rm fm$.}
	\label{fig:nofk_r0}
\end{figure}

Ground-state properties of light- and medium-mass nuclei, such as binding energies, charge radii, 
and charge form factors, are independent of the choice of the coordinate-space 
cutoff for the employed local chiral interactions~\cite{Lynn:2017,Lonardoni:2018prl,Lonardoni:2018prc}. 
However, the effect of using softer potentials (larger coordinate-space cutoff) is visible in the momentum 
distributions, as shown in~\cref{fig:nofk_r0}. 
For a given system, different interactions provide a similar description
of the mean-field part of the momentum distribution ($k\lesssim k_F$). Higher momentum 
components of $\rho_N(k)$ are instead reduced for softer potentials.
However, as already discussed above, for a given interaction the universality of the high-momentum 
components of $\rho_N(k)$ is preserved.

In \isotope[4]{He} both parametrizations of the three-body force ($E\tau$ and $E\mathbbm1$) for a given cutoff 
provide consistent results for the single-nucleon momentum distribution. The same observation applies to \isotope[16]{O}, 
for which, however, the $E\tau$ parametrization with cutoff $R_0=1.2\,\rm fm$ has not been considered 
in this work due to the large overbinding predicted by such a potential~\cite{Lonardoni:2018prl,Lonardoni:2018prc}.
Calculations for \isotope[12]{C} have been performed for the $E\tau$ $R_0=1.0\,\rm fm$ potential only
due to the large computational cost.

\begin{figure}[t]
	\includegraphics[width=\linewidth]{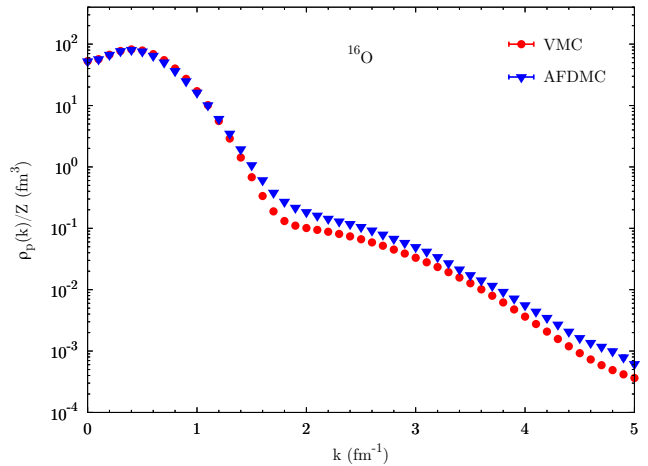}
	\caption[]{VMC and AFDMC proton momentum distributions in \isotope[16]{O}.}
	\label{fig:nofk_afdmc}
\end{figure}

\begin{figure*}[htb]
	\includegraphics[width=0.48\linewidth]{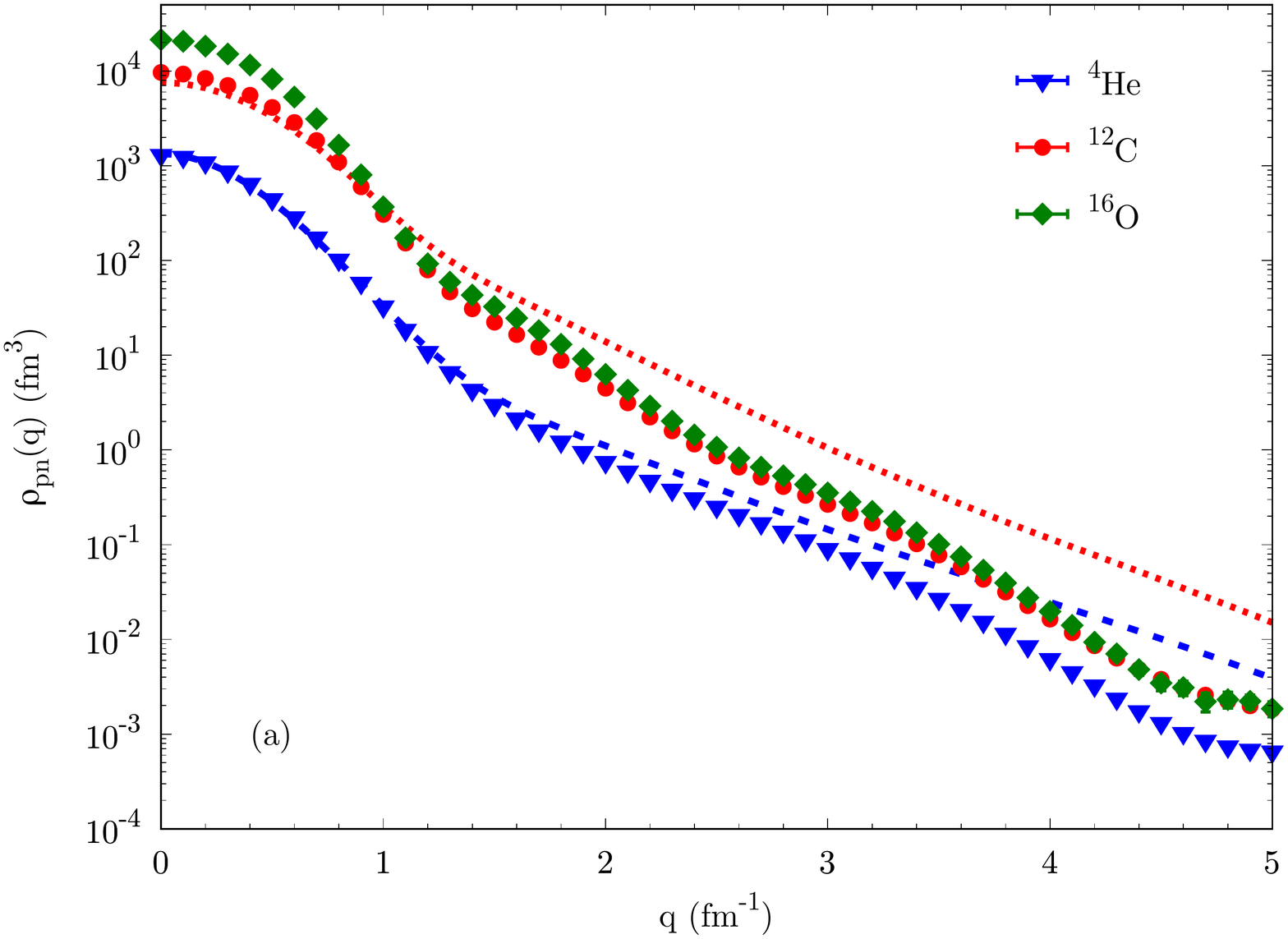}\qquad\includegraphics[width=0.48\linewidth]{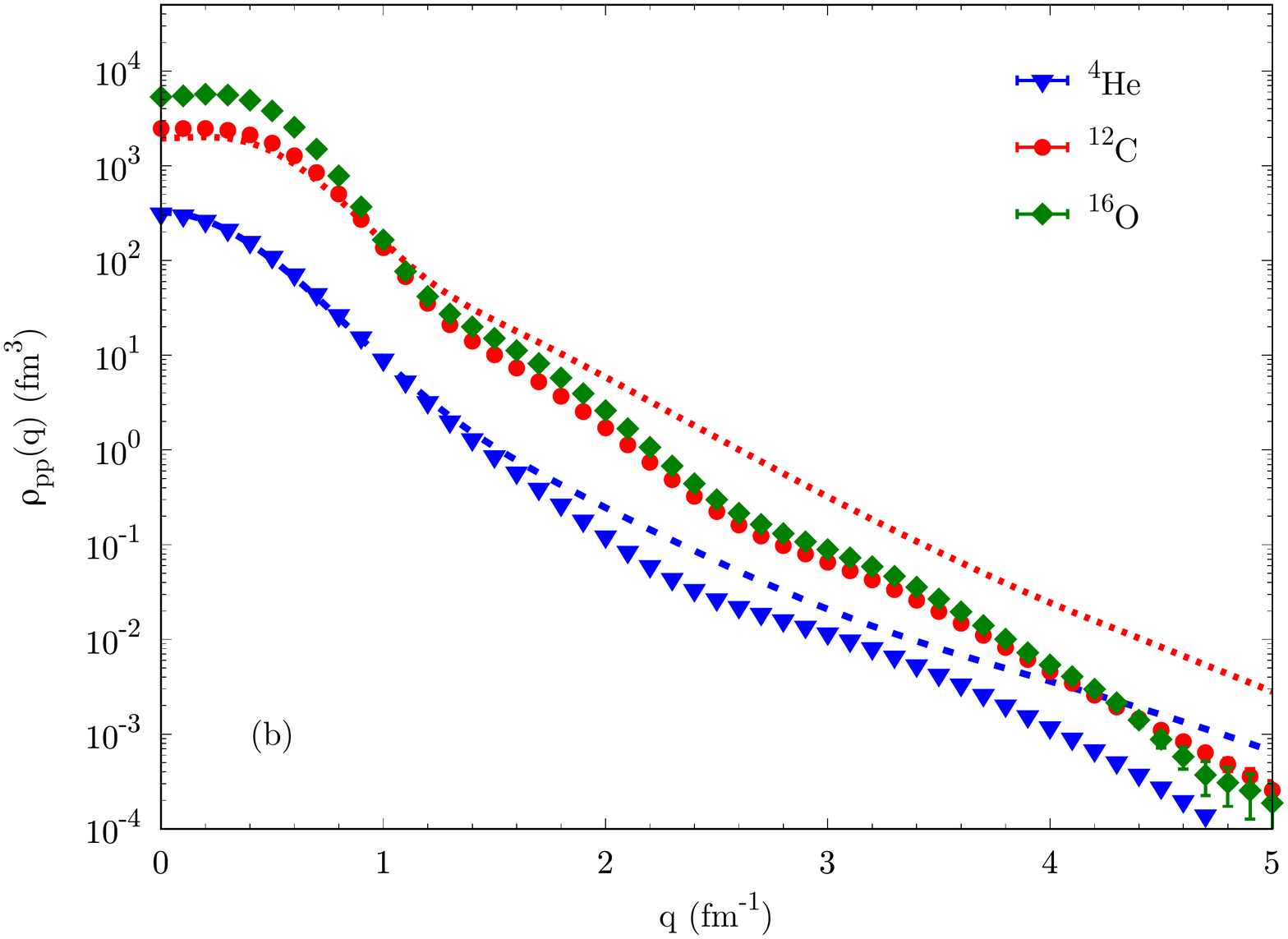}
	\caption[]{Two-nucleon momentum distributions integrated over $Q$:
		(a) $pn$ pairs and (b) $pp$ pairs. 
		Solid symbols are the results for the N$^2$LO $E\tau$ potential with cutoff $R_0=1.0\,\rm fm$.
		Lines are VMC results for the AV18+UX potential~\cite{Wiringa:2014,Wiringa:rhok12}.}
	\label{fig:nofq}
\end{figure*}

\begin{figure*}[htb]
	\includegraphics[width=0.48\linewidth]{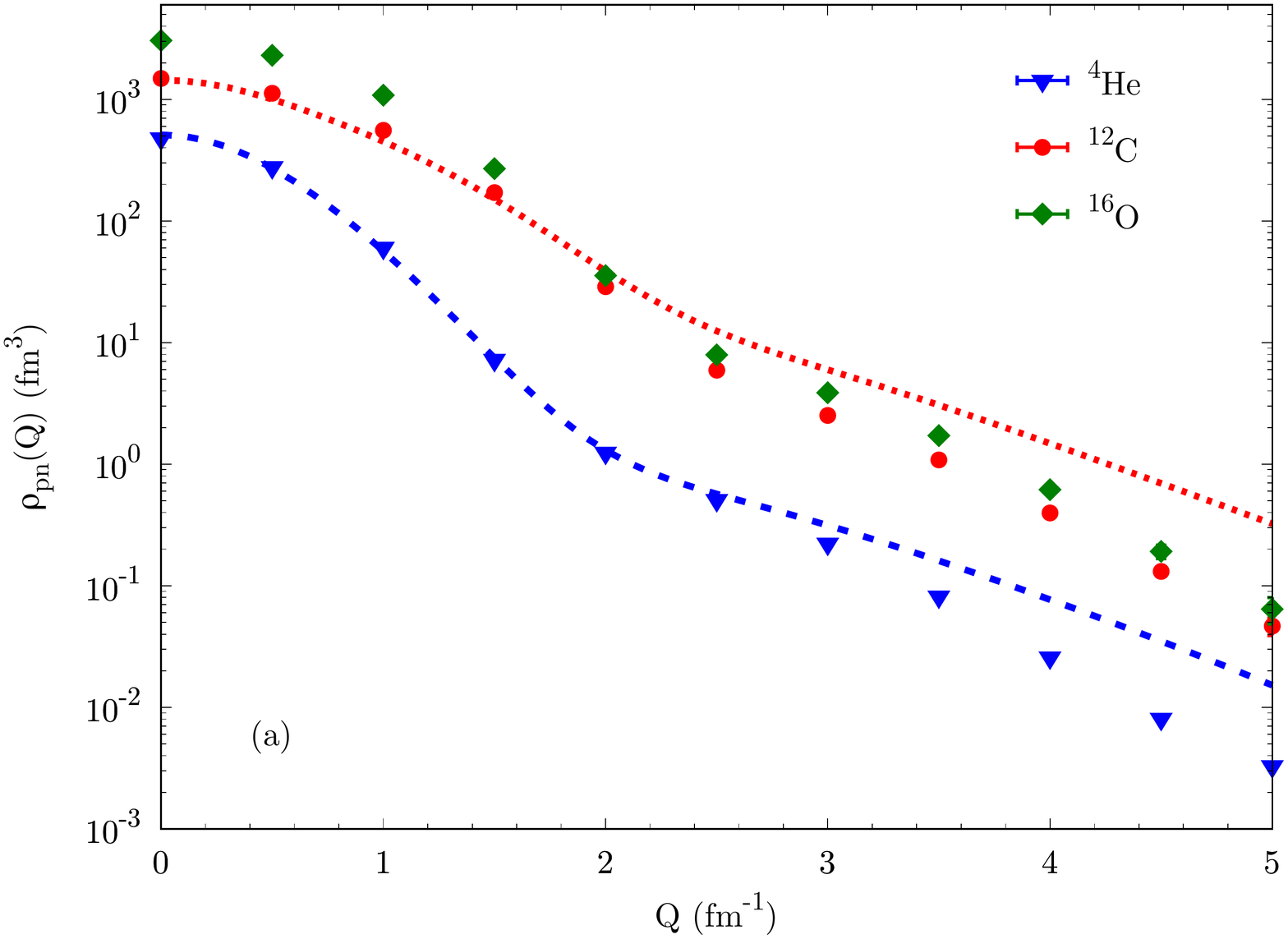}\qquad\includegraphics[width=0.48\linewidth]{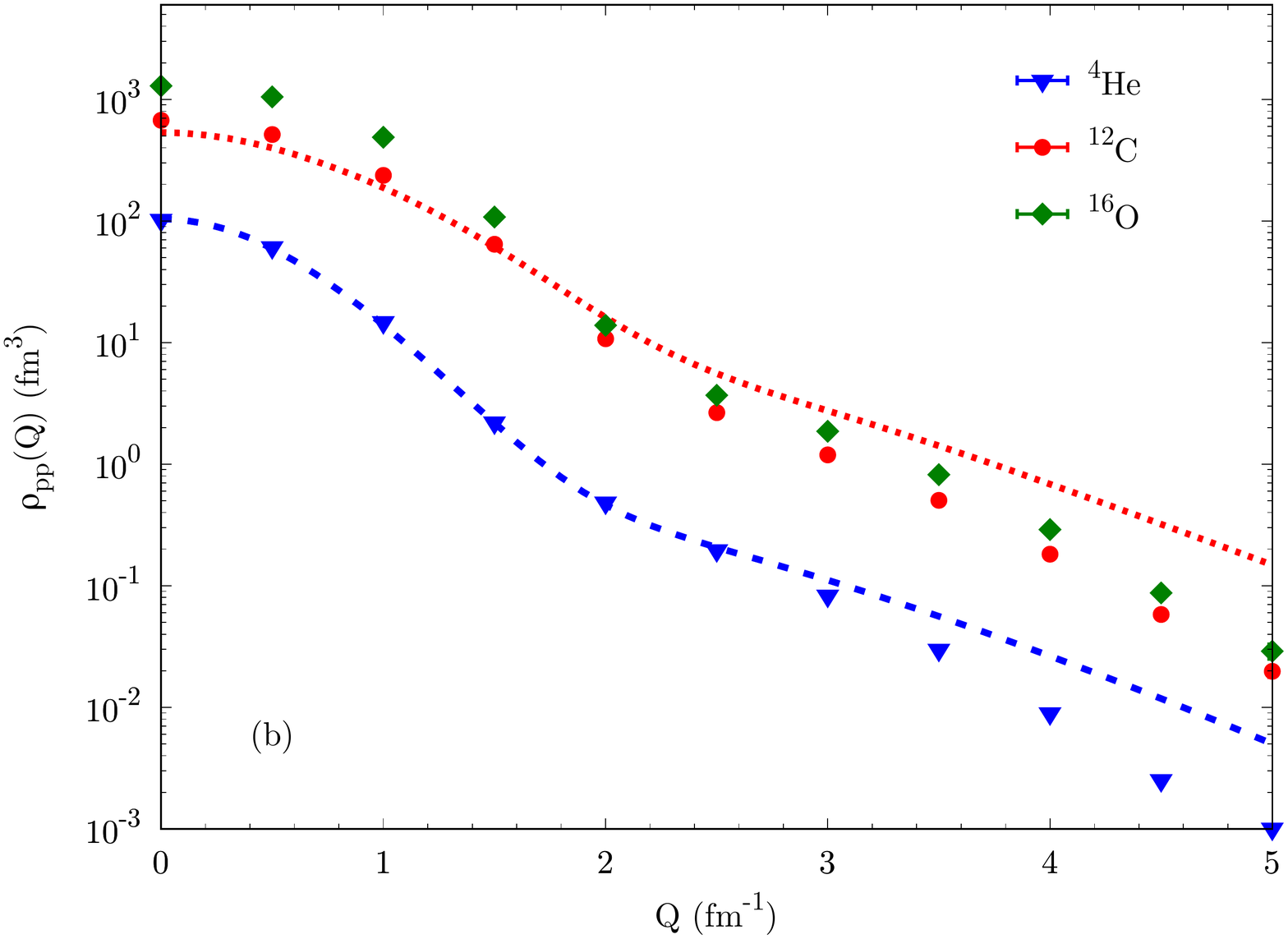}
	\caption[]{Two-nucleon momentum distributions integrated over $q$:
		(a) $pn$ pairs and (b) $pp$ pairs. 
		Solid symbols are the results for the N$^2$LO $E\tau$ potential with cutoff $R_0=1.0\,\rm fm$.
		Lines are VMC results for the AV18+UX potential~\cite{Wiringa:2014,Wiringa:rhok12}.}
	\label{fig:nofqq}
\end{figure*}

\Cref{fig:nofk_afdmc} shows the VMC and AFDMC results for the proton momentum distribution in \isotope[16]{O}, 
where the latter are extrapolated from mixed estimates (see Ref.~\cite{Lonardoni:2018prc} for details). 
The AFDMC results are expected to be more accurate, as they evaluate the
expectation values obtained through imaginary time propagation to the ground state. 
However, the AFDMC calculation for $A=16$ required $\approx 10^5$ more computing time than 
that of the VMC, due to the different scaling of the two Monte Carlo algorithms with the number of particles 
and the additional statistics required to obtain comparable statistical errors.
In \isotope[16]{O}, the AFDMC results are $\approx35\%$ higher than the VMC in the high-momentum 
region $(k\gtrsim 2\,\rm fm^{-1})$. Similar results are found for \isotope[4]{He}.
Improved trial wave functions, such as those described in Ref.~\cite{Lonardoni:2018prc}, 
could, in principle, bring the VMC results in closer agreement to those of AFDMC. However, 
the additional required computing time could be prohibitive for larger systems, already 
at the VMC level. Studies in this direction are in progress.

\section{Results: two-nucleon momentum distributions}
We present in~\cref{fig:nofq} the two-nucleon momentum distributions 
as a function of the relative momentum $q$ (the center-of-mass momentum $Q$ is integrated
over). Solid symbols are the results for the N$^2$LO $E\tau$ interaction with cutoff $R_0=1.0\,\rm fm$. 
Dotted and dashed lines refer to results employing phenomenological potentials, where available~\cite{Wiringa:2014,Wiringa:rhok12}. 
In~\cref{fig:nofq}{\color{blue}(a)} [\cref{fig:nofq}{\color{blue}(b)}] the momentum distributions for
$pn$ ($pp$) pairs are shown. As for the single-nucleon momentum distributions, up to $k\lesssim k_F$ there is little 
difference in the physical description of $\rho_{N\!N}(q)$ provided by chiral and phenomenological interactions.
Higher momentum components of $\rho_{N\!N}(q)$ are instead reduced for local chiral forces, in particular in heavier 
systems. At $q=2\,\rm fm^{-1}$ $97.3(2)\%$ [$98.6(1)\%$] of the $4\,pn$ ($1\,pp$) pairs are accounted for in \isotope[4]{He}.
These percentages are $98.8(7)\%$ [$99.1(3)\%$] for the $36\,pn$ ($15\,pp$) pairs in \isotope[12]{C} and
$99.1(8)\%$ [$99.4(4)\%$] for the $64\,pn$ ($28\,pp$) pairs in \isotope[16]{O}. 

An alternative way to look at two-nucleon momentum distributions is to integrate~\cref{eq:rhoqq} over all $q$,
leaving a function $\rho_{N\!N}(Q)$ of the center-of-mass momentum $Q$ only. In~\cref{fig:nofqq} we show $\rho_{N\!N}(Q)$
results for the N$^2$LO $E\tau$ potential with cutoff $R_0=1.0\,\rm fm$ (solid symbols) compared to available 
results for phenomenological potentials (dotted and dashed lines)~\cite{Wiringa:2014,Wiringa:rhok12}.
As in~\cref{fig:nofq}, \cref{fig:nofqq}{\color{blue}(a)} [\cref{fig:nofqq}{\color{blue}(b)}] 
reports $pn$ ($pp$) momentum distributions. As already observed in 
Ref.~\cite{Wiringa:2014} for lighter nuclei and phenomenological potentials, $\rho_{N\!N}(Q)$ for a given system 
has a smaller falloff at large momentum compared to $\rho_{N\!N}(q)$. The ratio of $pn$ to $pp$ pair is also subject 
to a smaller variation over the range of $Q$. The same conclusions hold for local chiral interactions up to \isotope[16]{O}, 
the results of which are similar to those of phenomenological potentials up to $Q\approx2\,\rm fm^{-1}$.

\begin{figure}[t]
	\includegraphics[width=\linewidth]{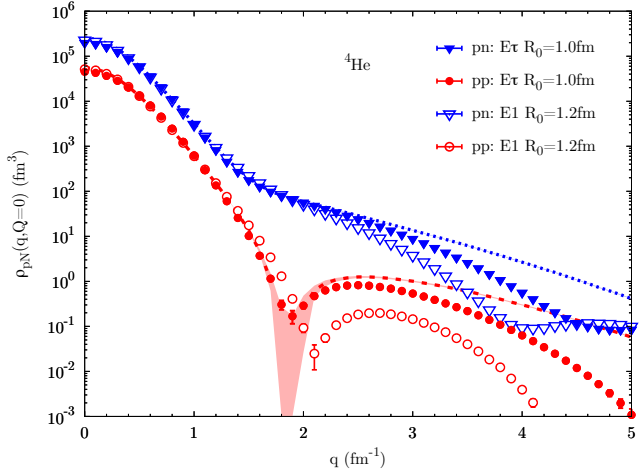}
	\caption[]{Two-nucleon momentum distributions as a function of $q$ for $Q=0$ in \isotope[4]{He}.
		Solid symbols are the results for the N$^2$LO $E\tau$ potential with cutoff $R_0=1.0\,\rm fm$.
		Empty symbols are the results for the N$^2$LO $E\mathbbm1$ potential with cutoff $R_0=1.2\,\rm fm$.
		Lines with error bands are VMC results for the AV18+UX potential~\cite{Wiringa:2014,Wiringa:rhok12}.}
	\label{fig:nofq+qq00_he4}
\end{figure}

\begin{figure}[b]
	\includegraphics[width=\linewidth]{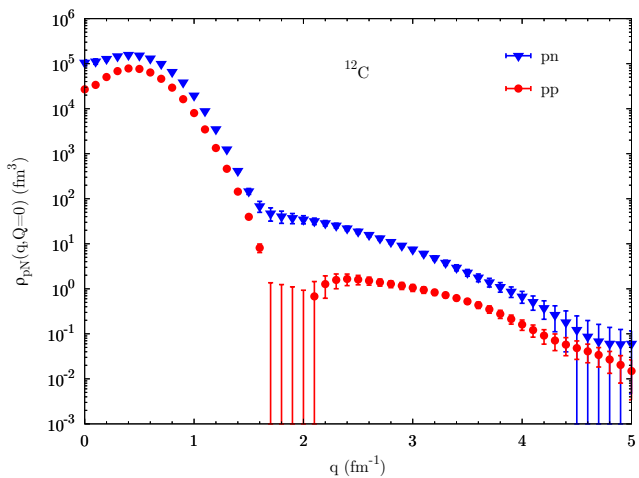}
	\caption[]{Two-nucleon momentum distributions as a function of $q$ for $Q=0$ in \isotope[12]{C}.
		The N$^2$LO $E\tau$ potential with cutoff $R_0=1.0\,\rm fm$ is used.}
	\label{fig:nofq+qq00_c12}
\end{figure}

\begin{figure}[t]
	\includegraphics[width=\linewidth]{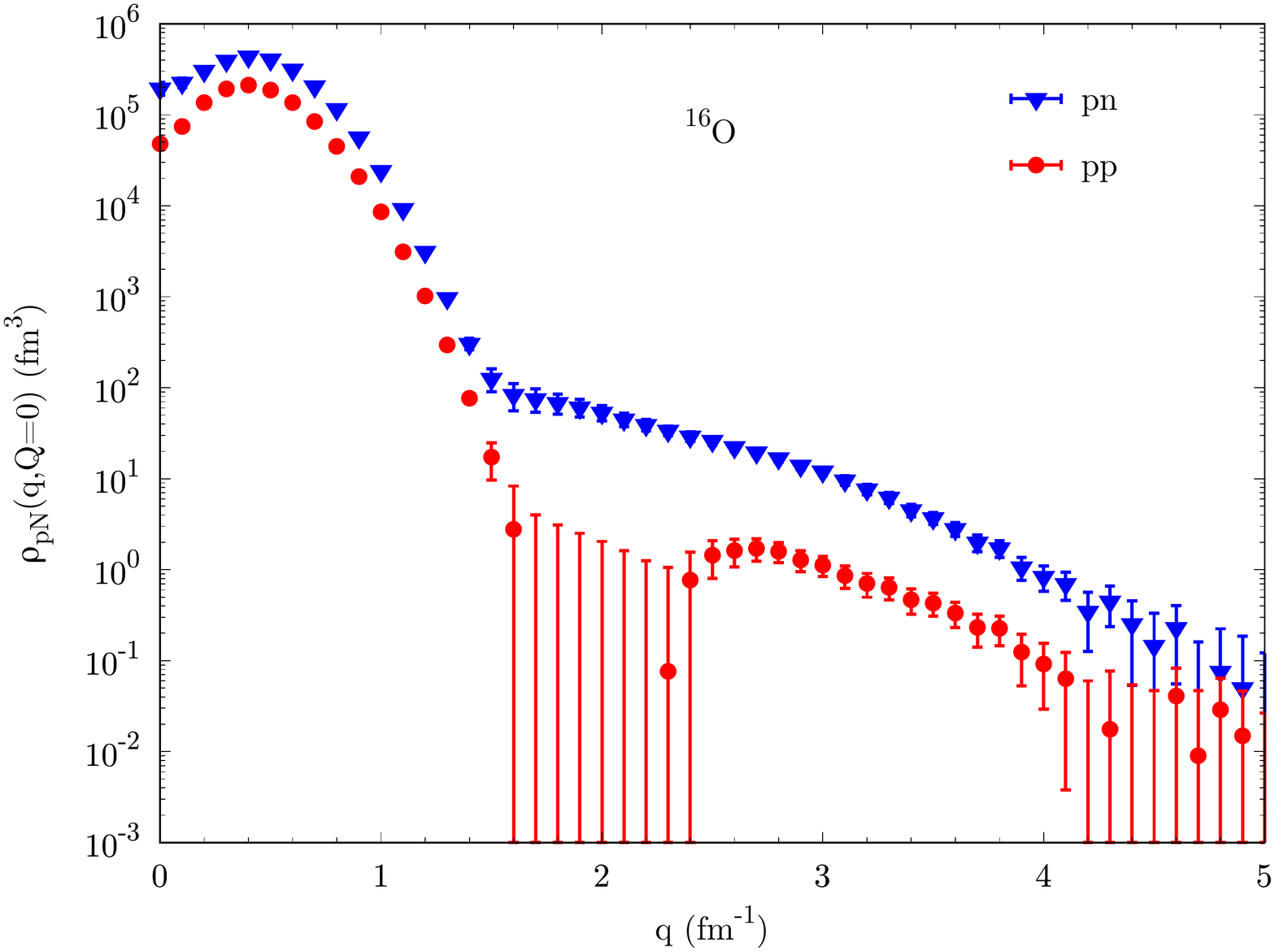}
	\caption[]{Same as~\cref{fig:nofq+qq00_c12} but for \isotope[16]{O}.}
	\label{fig:nofq+qq00_o16}
\end{figure}

The two-nucleon momentum distributions $\rho_{N\!N}(q,Q)$ at $Q=0$ (back-to-back pairs) in
\isotope[4]{He}, \isotope[12]{C}, and \isotope[16]{O} are shown 
in~\cref{fig:nofq+qq00_he4,fig:nofq+qq00_c12,fig:nofq+qq00_o16}, respectively.
Solid symbols are the results for the N$^2$LO $E\tau$ potential with cutoff $R_0=1.0\,\rm fm$.
Empty symbols are those for the N$^2$LO $E\mathbbm1$ potential with cutoff $R_0=1.2\,\rm fm$.
Blue triangles (red circles) indicate $pn$ ($pp$) pairs.
For $A=4$ the VMC results employing phenomenological potentials~\cite{Wiringa:2014,Wiringa:rhok12} 
are also reported for comparison. 
For \isotope[12]{C}, results are available for the harder interaction only.
In all systems $\rho_{N\!N}(q,Q=0)$ is larger for $pn$ pairs compared to $pp$
pairs, in particular for relative momentum in the range $q\approx1.5-2.5\,\rm fm^{-1}$. 
The $pp$ distributions present a node in this region, the position of which sits
around $2\,\rm fm^{-1}$ for all the nuclei considered in this work. 
$pn$ pairs show instead a deuteronlike distribution, with a change of slope
around $q=1.5\,\rm fm^{-1}$, as for phenomenological potentials~\cite{Wiringa:2014,Wiringa:rhok12}.
The ratio of $pn$ to $pp$ pairs in the region $q\approx1.5-2.5\,\rm fm^{-1}$ is $\gtrsim20$ in 
\isotope[4]{He} and $\gtrsim10$ in \isotope[12]{C} and \isotope[16]{O}. 
The same conclusion holds for both harder and softer potentials.
Although there are differences in the description of the two-nucleon
momentum distributions, the $pn$ to $pp$ ratio in the region $q\approx1.5-2.5\,\rm fm^{-1}$ is 
nearly independent of the employed local chiral interactions. 

\begin{figure}[t]
	\includegraphics[width=\linewidth]{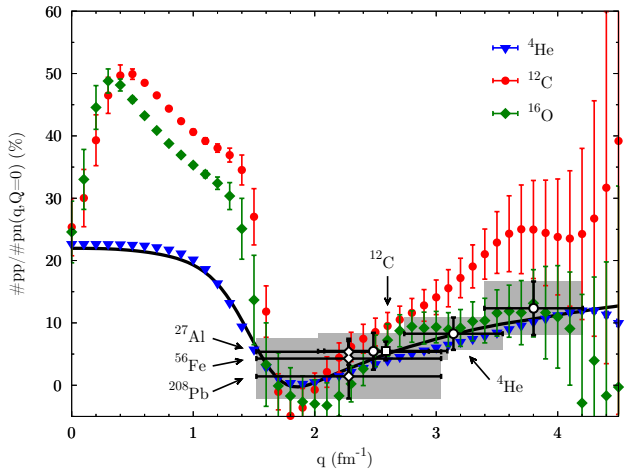}
	\caption[]{$pp$ pairs to $pn$ pairs ratio as a function of $q$ for $Q=0$.
		The N$^2$LO $E\tau$ potential with cutoff $R_0=1.0\,\rm fm$ is used.
		The solid curve was extracted from the two-body momentum distributions in \isotope[4]{He}
		for phenomenological potentials~\cite{Wiringa:rhok12}.
		Black empty symbols and gray bands were extracted from experimental data:
		\isotope[4]{He} from Ref.~\cite{Korover:2014}, \isotope[12]{C} from Ref.~\cite{Subedi:2008}, 
		and \isotope[27]{Al}, \isotope[56]{Fe}, and \isotope[208]{Pb} from Ref.~\cite{Hen:2014}.}
	\label{fig:nofq_qq00_pn}
\end{figure}

In~\cref{fig:nofq_qq00_pn} we report the ratio between $pp$ and $pn$ pairs as a function
of $q$ for back-to-back pairs. The N$^2$LO $E\tau$ potential with cutoff $R_0=1.0\,\rm fm$ is used.
Results for \isotope[4]{He} employing phenomenological potentials~\cite{Wiringa:rhok12} are shown 
for comparison (solid line).
Empty symbols are extracted from experimental data: circles for \isotope[4]{He} from Ref.~\cite{Korover:2014}, 
squares for \isotope[12]{C} from Ref.~\cite{Subedi:2008}, and diamonds for \isotope[27]{Al}, 
\isotope[56]{Fe}, and \isotope[208]{Pb} from Ref.~\cite{Hen:2014}.
For the employed local chiral interactions all nuclei are consistent with high-momentum
data extracted from experiments. 

Note that the wave function of~\cref{eq:psi} only includes linear spin/isospin-dependent 
two-body correlations; i.e., only one nucleon pair is correlated at a time. Improved two-body 
correlations (see Ref.~\cite{Lonardoni:2018prc} for details) are under study, but the 
increased computing time requested to evaluate the full wave function will make the
calculation of two-body momentum distributions for medium-mass nuclei computationally
challenging. However, preliminary tests in \isotope[4]{He} show an $\approx8-18\%$ 
variation of the $pp$ to $pn$ ratio in the range $q\approx2.5-4.0\,\rm fm^{-1}$, 
result still compatible with the available data extracted from experiments.

\begin{figure}[b]
	\includegraphics[width=\linewidth]{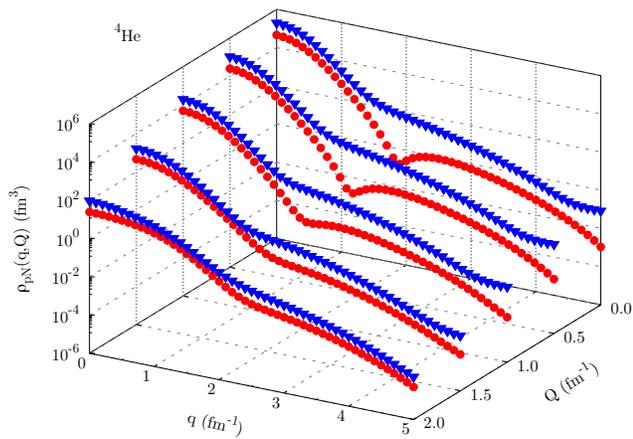}
	\caption[]{Two-nucleon momentum distributions in \isotope[4]{He} 
		for the N$^2$LO $E\tau$ potential with cutoff $R_0=1.0\,\rm fm$.
		Blue triangles refer to $pn$ pairs, and red circles refer to $pp$ pairs.}
	\label{fig:nofq+qq_he4}
\end{figure}

The electron scattering experiments necessarily involve two-nucleon currents,
which are not included in this work. Here we provide comparisons to
other calculations of single- and two-nucleon momentum 
distributions~\cite{Alvioli:2013,Wiringa:2014,Alvioli:2016,Lonardoni:2017}.
These currents provide a $30-40\%$ constructive interference in the
inclusive transverse quasielastic
electron scattering~\cite{Lovato:2016} and in the axial response relevant
to neutrino scattering~\cite{Lovato:2018}. It remains to be investigated
how they impact the back-to-back exclusive measurements.

\begin{figure}[t]
	\includegraphics[width=\linewidth]{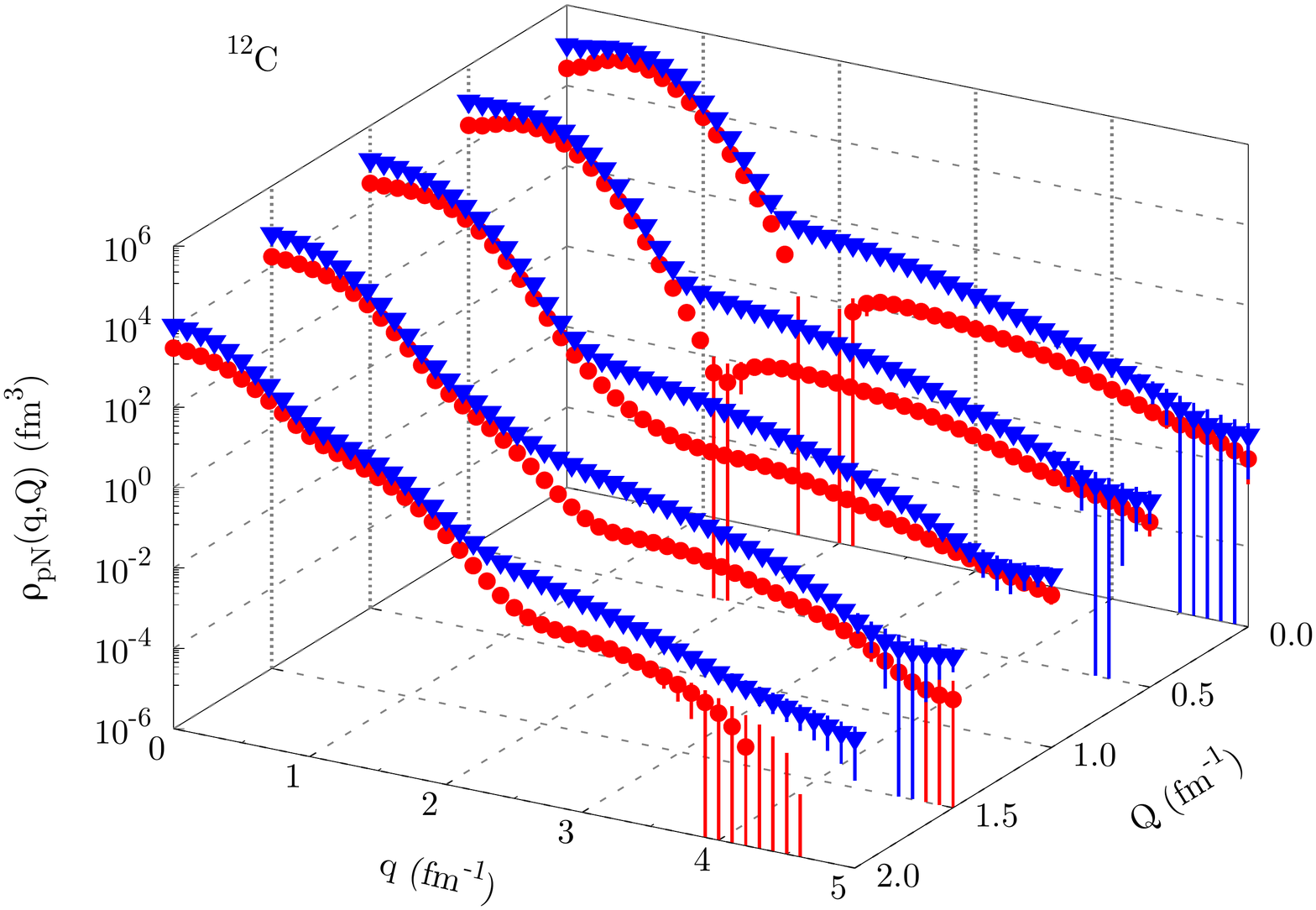}
	\caption[]{Same as~\cref{fig:nofq+qq_he4} but for \isotope[12]{C}.}
	\label{fig:nofq+qq_c12}
\end{figure}

\begin{figure}[b]
	\includegraphics[width=\linewidth]{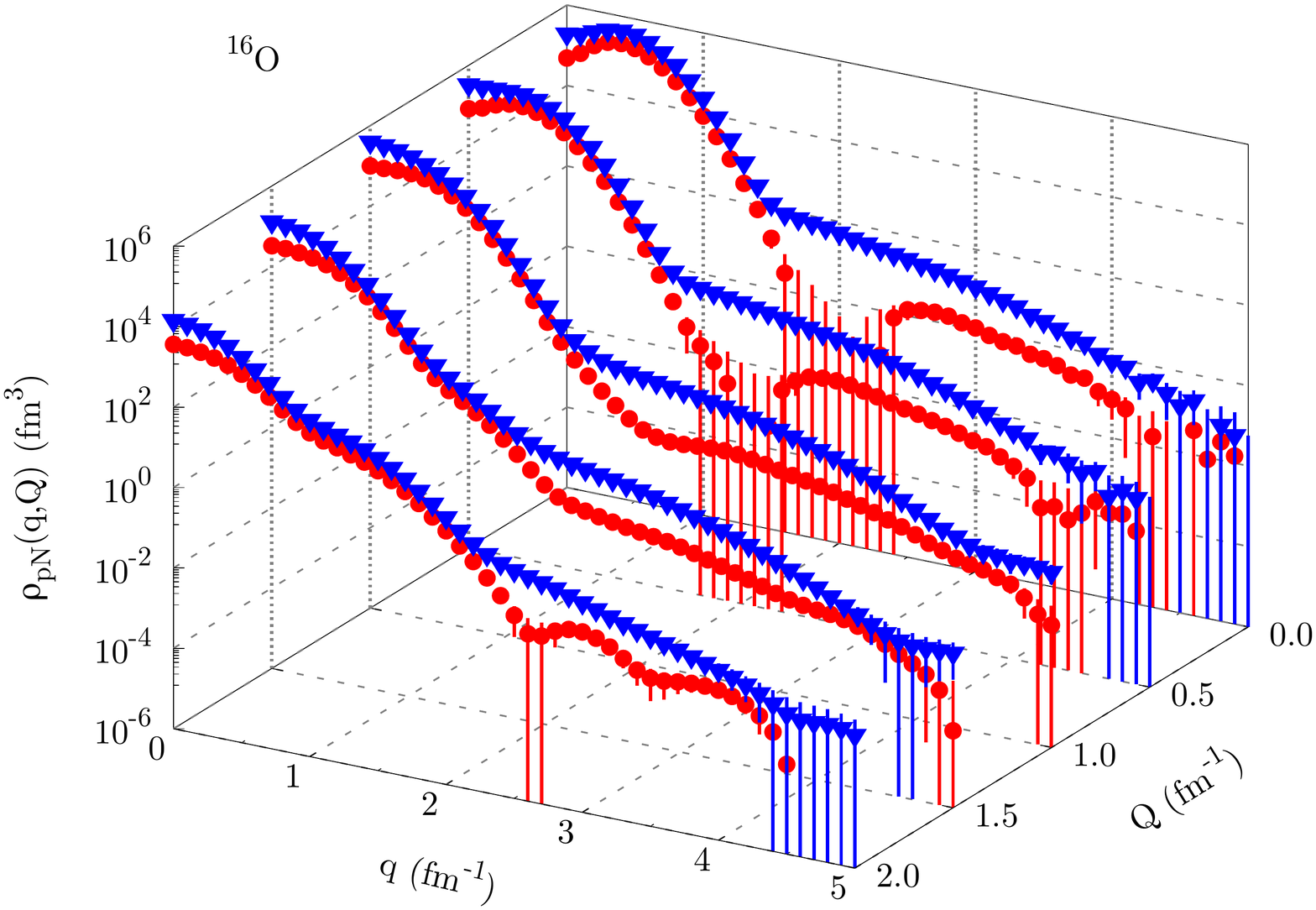}
	\caption[]{Same as~\cref{fig:nofq+qq_he4} but for \isotope[16]{O}.}
	\label{fig:nofq+qq_o16}
\end{figure}

Finally, the evolution of the two-nucleon momentum distribution as a function of $Q$ is
shown in~\cref{fig:nofq+qq_he4,fig:nofq+qq_c12,fig:nofq+qq_o16} for 
\isotope[4]{He}, \isotope[12]{C}, and \isotope[16]{O}, respectively.
As in the previous plots, blue triangles (red circles) are the results for 
$pn$ ($pp$) pairs employing the N$^2$LO $E\tau$ potential with cutoff $R_0=1.0\,\rm fm$.
The description of $\rho_{pn}(q,Q)$ and $\rho_{pp}(q,Q)$ in \isotope[4]{He} as $Q$ 
increases is analogous to that provided by phenomenological 
potentials~\cite{Wiringa:2014,Wiringa:rhok12}. The node in the $pp$ distribution
gradually disappears, while the deuteronlike distribution of $pn$ pairs is maintained
up to large $Q$. The same physical picture holds for larger
nuclei up to $A=16$. For $Q\gtrsim1.5\,\rm fm^{-1}$ the node in the $pp$ momentum
distributions is completely filled in, and the $pn$ to $pp$ ratio is largely reduced.

The tables of single- and two-nucleon momentum distributions in
\isotope[4]{He}, \isotope[12]{C}, and \isotope[16]{O} for local chiral potentials
are available as Supplemental Material~\cite{supp:2018}
and as part of the online quantum Monte Carlo momentum distribution 
collection~\cite{Wiringa:rhok,Wiringa:rhok12}.

\section{Summary}
We presented VMC calculations of the single- and two-nucleon 
momentum distributions in \isotope[4]{He}, \isotope[12]{C},
and \isotope[16]{O} employing local chiral interactions at N$^2$LO.
The description of the momentum distributions at low and
moderate momenta up to $\approx 2 k_F$ is similar to
that provided by phenomenological potentials at low momentum, while
higher-momentum components are typically reduced, consistent with 
the lower-energy regime of chiral EFT interactions.

The effect of short-range correlations on the high-momentum 
components of the single-nucleon momentum distribution is found to be large
and dominant also for local chiral interactions. The universality of the tail of the 
momentum distribution is confirmed, but only within the same family of interactions. 

The two-nucleon momentum distributions as a function of the relative
momentum $q$ of the nucleon pair, of the center-of-mass momentum $Q$ of the pair, 
and of both $q$ and $Q$ are shown. The results for back-to-back
pairs confirm the large $pn$ to $pp$ pairs ratio in the regime 
$q\approx1.5-2.5\,\rm fm^{-1}$ up to \isotope[16]{O}, which appears to be
independent of the employed interaction scheme. 
The $pp$ to $pn$ ratio for local chiral interactions is compatible with 
available experimental data extracted from electron scattering experiments
in the range $q\approx2.5-4.0\,\rm fm^{-1}$ up to $A=16$.

It will be interesting to analyze the results of this work using factorized 
asymptotic wave functions and the short-range correlations as done in Ref.~\cite{Weiss:2018}
for phenomenological potentials. This will provide information about how sensitive 
are the contacts and ratios of contacts to the scale and scheme of the calculations,
opening the possibility of relating a very large class of observables to ground-state 
calculations.

\acknowledgments{We thank R.~B.~Wiringa and O.~Hen for many valuable discussions.
The work of D.L. was supported by the U.S. Department of Energy,
Office of Science, Office of Nuclear Physics, under Contract 
No. DE-SC0013617, and by the NUCLEI SciDAC program.
The work of S.G. and J.C. was supported by the NUCLEI SciDAC program,
by the U.S. Department of Energy, Office of Science, Office of Nuclear
Physics, under Contract No. DE-AC52-06NA25396, and by the LDRD program
at LANL.
X.B.W. is grateful for the hospitality and financial support of LANL and the
National Natural Science Foundation of China under Grants No. U1732138, 
No. 11505056, No. 11605054, and No. 11747312, and China Scholarship Council 
(Grant No. 201508330016).
Computational resources have been provided by Los Alamos Open
Supercomputing via the Institutional Computing (IC) program and by the
National Energy Research Scientific Computing Center (NERSC), which is
supported by the U.S. Department of Energy, Office of Science, under
Contract No. DE-AC02-05CH11231.}

\end{document}